# Temperature dependence of anisotropic thermal conductivity tensor of bulk black phosphorus

Bo Sun, [1] Xiaokun Gu, [2] Qingsheng Zeng, [3] Xi Huang, [1] Yuexiang Yan, [1] Zheng Liu, [3] Ronggui Yang [2] and Yee Kan Koh [1, 4, *]

1  Department of Mechanical Engineering, National University of Singapore, 117576, Singapore

2  Department of Mechanical Engineering, University of Colorado at Boulder, CO 80309, USA

3 Centre for Programmed Materials, School of Materials Science and Engineering, Nanyang   Technological University, 639798, Singapore

* mpekyk@nus.edu.sg

Thermal transport in layered, two-dimensional (2D) black phosphorus (BP) is of great interest, not only due to its importance in the designs of BP devices,[1] but also because it provides a unique platform to study the physics of heat transport in highly anisotropic materials.[2] BP belongs to the orthorhombic Cmca point group,[3] with its puckered honeycomb basal planes weakly bonded together by interlayer van der Waals' forces. Due to the nature of its crystal structure, second order tensors (e.g., the thermal conductivity tensor $\Lambda$) of BP have three independent components along the principal axes of zigzag (ZZ), armchair (AC) and through-plane (TP), see **Figure 1a**, and the thermal conductivity tensor is strongly anisotropic along these axes.[4] (In this paper, we use $\Lambda_{ZZ}$, $\Lambda_{AC}$ and $\Lambda_{TP}$ to denote the three independent components of the thermal conductivity tensor.) Here, we accurately measured and report the anisotropic thermal conductivity tensor ($\Lambda_{ZZ}$, $\Lambda_{AC}$ and $\Lambda_{TP}$) of bulk BP in a temperature range of $80 \leq T \leq 300$ K. Our temperature dependence measurements provide a crucial benchmark for future studies of anisotropic heat transport in BP and phosphorene.

To date, there are only few experimental works on anisotropic thermal conductivity of BP, even at 300 K. Luo *et al.*[5] and Lee *et al.*[6] measured BP flakes with a thickness of 9 – 30 nm and 60 – 310 nm using the opto-thermal Raman method and the micro-bridge technique, respectively, and reported $\Lambda_{ZZ} = 11 – 45$ W m$^{-1}$ K$^{-1}$ and $\Lambda_{AC} = 5 – 22$ W m$^{-1}$ K$^{-1}$ at room temperature. These values of $\Lambda_{ZZ}$ and $\Lambda_{AC}$ are substantially lower than predictions by first-principles calculations[4, 7, 8] for bulk BP and phosphorene. While these low values of thermal conductivity were attributed to additional boundary scattering of phonons in the thin flakes,[5, 6] we note that scattering of phonons along the basal planes by the interfaces is rather weak[9] and thus this explanation might not be satisfactory. The low values could also originate from degradation of the BP flakes by oxidation,[10] as the BP flakes in both studies were exposed to the air for a substantial amount of time during sample preparation and measurements. With the degradation, the reported thermal conductivity is probably not intrinsic. Jang et al.[11] encapsulated their BP flakes of thickness of 138 – 552 nm with a 3-nm $Al_2O_3$ overlayer, and obtained $\Lambda_{ZZ}$ = 63 – 86 W m$^{-1}$ K$^{-1}$, $\Lambda_{AC}$ = 26 – 34 W m$^{-1}$ K$^{-1}$ and $\Lambda_{TP}$ = 3.2 – 4.5 W m$^{-1}$ K$^{-1}$ from their time-domain thermoreflectance (TDTR) measurements. Zhu et al.[4] derived $\Lambda_{ZZ}$ = 84 - 101 W m$^{-1}$ K$^{-1}$, $\Lambda_{AC}$ = 26 - 36 W m$^{-1}$ K$^{-1}$ and $\Lambda_{TP}$ = 4.3 - 5.5 W m$^{-1}$ K$^{-1}$ from their measurements on BP of thickness of 30 – 50 μm using time-resolved magneto-optic Kerr effect (TR-MOKE). Although Jang et al.'s and Zhu et al.'s samples were not seriously oxidized, their pump-probe measurements in the through-plane direction might be lower than the intrinsic $\Lambda_{TP}$ because the mean-free-paths ($\ell$) of a substantial portion of heat-carrying phonons are much longer than the characteristic length scales of their measurements (<500 nm), *i.e.*, the thickness of the samples or the thermal penetration depth *d*.[12-14] In fact, we obtained a $\Lambda_{TP}$ value ~25 % higher than Jang et al.'s and Zhu et al.'s measurements,[4, 11] when we used a much lower modulation frequency in our measurements to achieve a larger thermal penetration depth.

With the relatively few published works on the thermal properties of BP, knowledge of anisotropic heat transport in BP and other layered materials (e.g., the mean-free-paths of phonons in the through-plane and in-plane directions) is still incomplete. Even for the most well-studied graphite, the consensus has yet been achieved. For example, while Zhang et al.[15] reported that the through-plane thermal conductivity converges when the thickness of graphite film is above 500 nm, Fu et al.[16] found that the thermal conductivity of a 700 nm thin film is still 27% lower than the bulk value. In this work, we employ frequency-dependent TDTR measurements[12, 17] and the first-principles calculations to determine the mean-free-paths distribution of phonons in bulk BP along the through-plane direction. We find that at 300 K the span of phonon mean-free-paths along the TP axis is surprisingly broad from sub-10 nm to 1 μm. Also, we find that frequency-dependent phonon relaxation times are mostly isotropic in directions when both the vibrations and the propagation directions of phonons are in-plane, but are significantly suppressed when either the vibrations or the propagation directions of phonons are out-of-plane.

Our samples are three millimeter-sized bulk BP samples, see **Figure 1b**. The largest sample, denoted by BP-1, with a dimension of ~8 mm × 5 mm × 0.3 mm, was purchased from HQ Graphene. The other two smaller samples, BP-2 and BP-3, with a dimension of 5 mm × 1.5 mm × 0.1 mm and 5 mm × 1 mm × 0.1 mm, were home-grown. All three BP samples were grown by a mineralizer-assisted gas-phase transformation method,[18] through a short way transport reaction of red phosphorus with $Sn/SnI_4$ as the mineralization additive. Sizes of several millimeters to centimeters are realized by controlling the cooling speed during the growth process. Details of our sample preparation are presented in the Experimental Section. To prepare the samples for measurements, we mounted BP-2 and BP-3 on Si substrates for easier handling. (BP-1 is exempted from this step since it is large and thick enough.) We then exfoliated the top layers of the BP samples to expose the fresh BP surfaces, and immediately loaded the samples into an ultra high vacuum (UHV) thermal evaporation chamber for deposition of a 100 nm thick Al film. The Al film acts as the transducer for our measurements, and also passivates the fresh BP surfaces. We ensure that the total time that the BP samples were exposed to air is less than 1 minute to minimize oxidation of BP.

Before the thermal measurements, we identify the crystallographic orientations of the BP samples by polarized Raman spectroscopy,[19] see **Figure 1c**. To improve the accuracy of our determination of the crystallographic orientations of BP, we conduct the polarized Raman measurements for every 15°, and fit the measurements with a $\sin^2(2\theta)$ function to determine ZZ and AC axes, see **Figure 1d** and the Experimental Section. We also ensure that the basal planes are perpendicular to the sample surface via x-ray diffraction (XRD) scan of the BP-1 sample. In the XRD spectrum, we observed only (0 n 0) peaks, where n = 4, 6, and 8, as expected, see **Figure 1e**.

We measure the through-plane thermal conductivity $\Lambda_{TP}$ of the BP samples by the frequency-dependent time-domain thermoreflectance (TDTR),[12, 20] and the two in-plane thermal conductivities $\Lambda_{ZZ}$ and $\Lambda_{AC}$ by the beam-offset TDTR method.[21] Details of our implementation, data analysis and signal processing of TDTR and beam-offset TDTR methods are presented in the Experimental Section and the Supporting Information. In both methods, two parameters must be carefully considered: the $1/e^2$ radii $w_0$ of the laser beams and the modulation frequency $f$ of the pump beam. (The modulation frequency $f$ affects heat diffusion during the measurements, which is characterized by a parameter called the thermal penetration depth $d$; $d = \sqrt{\Lambda/\pi C f}$, where $\Lambda$ is thermal conductivity and $C$ is volumetric heat capacity.) For accurate measurements of $\Lambda_{TP}$, we employed a large $w_0$ (25 μm) and a low $f$ (0.5 MHz), and ensure that heat diffusion is mainly one-dimensional in the through-plane direction. The low $f$ is crucial to ensure that our TDTR measurements approach the intrinsic $\Lambda_{TP}$, since TDTR is not sensitive to heat transport by ballistic phonons with mean-free-path $\ell \gg 2d$.[12, 13, 22] (Interpretation of the frequency dependence of TDTR measurements is discussed below.) For beam-offset TDTR method, we employed a low $f = 0.5$ MHz and a small $w_0$ (as small as 2.5 μm), to ensure that heat mainly spreads along the basal planes and is predominantly determined by $\Lambda_{ZZ}$ and $\Lambda_{AC}$. At $f = 0.5$ MHz and $T = 80$ K, heat diffuses up to 50 μm in either side of the pump beam, see Figure 2c. Thus, for accurate measurements of the intrinsic thermal conductivity tensor of bulk BP at 80 K, the lateral dimension of the BP samples should be >100 μm.

To ensure the accuracy and consistency of our beam-offset measurements, we systematically performed multiple measurements on our BP samples: 1) we rotated our BP sample by 90° and performed the beam-offset measurements when the ZZ axis of the sample was horizontal and when it was vertical, relative to the optical table. We obtained identical results. 2) We performed an additional measurement 6 months after the initial sample preparation, to check for any degradation of our BP samples. No observable degradation was found. 3) We rotated our samples by ±30° and performed measurements along different crystallographic orientations. We achieved excellent agreement for all these measurements, see **Figure 2** and more detailed discussion in the Experimental Section.

We performed multiple measurements on 3 BP samples and obtained similar values for $\Lambda_{ZZ}$, $\Lambda_{AC}$ and $\Lambda_{TP}$, see **Table 1** for a summary of the average values of our measurements performed using different modulation frequency $f$ and laser spot size $w_0$ at room temperature. On average, we obtained $\Lambda_{ZZ} = 83 \pm 10$ W m$^{-1}$ K$^{-1}$, $\Lambda_{AC} = 28 \pm 5$ W m$^{-1}$ K$^{-1}$, and $\Lambda_{TP} = 6.5 \pm 0.8$ W m$^{-1}$ K$^{-1}$ at room temperature. For $\Lambda_{ZZ}$ and $\Lambda_{AC}$, our values are similar to the values obtained by Zhu et al.[4] and Jang et al.[11] For $\Lambda_{TP}$, our value is >25% higher than the prior reported $\Lambda_{TP}$ measured at a higher $f$ of 9 – 10 MHz.[4, 11] We believe that prior $\Lambda_{TP}$ is artificially low because with the high $f$, a substantial portion of phonons remain nonequilibrium within a region of $2d \approx 600$ nm and heat transport by these nonequilibrium phonons are not registered in TDTR measurements, see the discussion below on the frequency dependence of TDTR measurements. We note that we derived an apparent through-plane thermal conductivity of 5.4 W m$^{-1}$ K$^{-1}$ when we applied $f = 10$ MHz, similar to what Zhu *et al.* reported.[4] In our measurements, the diameters of our laser beams vary from 5 µm to 50 µm, much larger than the mean-free-paths of phonons in BP (see Figure 5b and Figure S11 in the Supporting Information). Thus, our results do not depend on the size of the laser beams, see Table 1. Our values thus represent the intrinsic thermal conductivity tensor of bulk BP.

We compare the derived $\Lambda_{TP}$, $\Lambda_{ZZ}$ and $\Lambda_{AC}$ in the temperature range of 80 – 300 K to first-principles calculations of the thermal conductivity of bulk BP by Zhu *et al.*[4] and of phosphorene by Jain *et al.*,[7] Zhu *et al.*[4] and Li *et al.*[8] in **Figure 3**. We find that within the temperature range, all three components of the measured thermal conductivity tensor are inversely proportional to $T$, which agrees well with the predictions from the first-principles calculations. The $T^{-1}$ dependence suggests that, in all crystallographic orientations, heat is carried mainly by phonons that are all predominantly scattered by Umklapp processes, despite the anisotropy that we observed in the thermal conductivity. Our measured $\Lambda_{AC}$ and $\Lambda_{TP}$ are similar to the first-principle calculations of bulk BP by Zhu *et al.*,[4] while the measured $\Lambda_{ZZ}$ are ~20 % lower than the first-principle prediction. Comparing to phosphorene, our $\Lambda_{ZZ}$ ($\Lambda_{AC}$) are 30 % (9 %) lower than the first-principles predictions by Jain *et al.*[7] The difference could be attributed to reduction in $\Lambda_{ZZ}$ and $\Lambda_{AC}$ of bulk BP due to enhanced scattering of out-of-plane TA1 phonons in bulk BP compared to that in phosphorene, similar to enhanced scattering of TA1 phonons in graphite compared to that in graphene.[23] (We used TA1 and TA2 to denote the lower and higher energy branches of the transverse phonons, respectively, see the phonon dispersion in Figure S8 in the Supporting Information. Along the basal planes (e.g., ZZ and AC), TA1 and TA2 refers to out-of-plane ZA and in-plane vibration modes.)

We present the thermal conductance ($G$) of Al/BP interfaces in **Figure 4**. From the temperature dependence, we confirm that heat is mainly carried by phonons across the BP interface. At 300 K, we find that $G = 72$ MW m$^{-2}$ K$^{-1}$; this value of $G$ is stable over a period of >6 months, indicating that our BP samples did not degrade over time. $G$ of Al/BP is nearly twice of $G$ of NbV/AlO$_x$/BP[11] and TbFe/BP[4], and is significantly higher than that of Al/graphite[24]. We attribute this observation to a better match of the Debye temperature between Al and BP[25], compared to Al and graphite.

To further understand the physics of anisotropic heat transport in BP, we employ the frequency-dependent TDTR measurements[12, 17] and first-principles-based phonon Boltzmann transport equation (BTE) calculations,[26] two powerful tools that were recently developed to study phonon mean-free-paths. Frequency dependence of TDTR measurements has been developed into a convenient approach to probe the mean-free-paths of phonons,[12, 17] even though questions on how to accurately interpret the measurements (e.g., how to handle the nonequilibrium heat transfer across the interfaces[27]) still remain. In this work, we use a simple approximation, presented below, to analyze our frequency-dependent TDTR measurements. We do not employ the more sophisticated approaches such as solving the BTE on the sample geometry[13] or reconstruction of phonon mean-free-paths using a complex optimization procedure[28], because 1) the main source of uncertainty in modeling is how to handle the transmission, reflection and scattering of phonons at the interface, which is not mitigated in these more sophisticated approaches; and 2) our simple analysis below might underestimate, if any, the distribution of phonon mean-free-paths. Thus, the conclusion that we derived from our frequency-dependent measurements – the distribution of the mean-free-paths of phonons along the through-plane direction is rather broad – should remain valid even if the more sophisticated analyses are applied.

Thus, we instead rely on the conclusions that we derived from our previous solution of the BTE on a semi-infinite solid using a boundary condition relevant to TDTR and FDTR. Our previous BTE calculations showed that as long as the distribution of phonon mean-free-path is sufficiently wide (e.g., in alloys when heat is carried mainly by low-energy phonons[27]), the apparent thermal conductivity measured by TDTR could be crudely approximated by assuming an additional boundary scattering at a characteristic length of $L_c = 2d$.[13] Since the phonon mean-free-paths in BP span more than two orders of magnitude according to our first-principles calculations, see below, the use of the criterion $L_c = 2d$ should be acceptable. We emphasize again that the thermal conductivity accumulation function derived using this simplified approach only gives a crude approximation to the distribution of phonon mean-free-paths in BP.

We first compare the apparent through-plane thermal conductivity $\Lambda_{TP}$ that we derived from our frequency-dependent TDTR measurements with a modulation frequency of $0.5 \leq f \leq 10$ MHz, to prior measurements on BP, see Figure 5a. We plot our frequency dependent $\Lambda_{TP}$ at 80 K, 150 K and 300 K as a function of $L_c = 2d$. We observe a considerable $L_c$ dependence for the apparent $\Lambda_{TP}$ at 300 K; $\Lambda_{TP}$ does not converged even when $L_c = 1$ µm. From the $L_c$ dependence, we expect that the phonon mean-free-paths along through-plane direction are relatively long, although the thermal conductivity is low.

In the same figure, we also plot prior measurements of $\Lambda_{TP}$ of BP flakes, as a function of either *2d* or film thickness *h*, whichever smaller. While we find that prior measurements on BP flakes by Jang *et al.*[11] and Zhu *et al.*[4] generally agree with our frequency-dependent measurements, we notice that two data points from Jang *et al.* significantly deviate, see Figure 5a. We attribute the discrepancy to the higher-than-usual level of uncertainty for measurements by Jang *et al*. Unlike the standard analysis of our TDTR measurements with only two unknowns (the thermal conductance of Al/BP interface and $\Lambda_{TP}$ of BP), there are four unknowns in the analysis of Jang *et al.*'s measurements including the thermal conductance of NbV/BP interfaces, the thickness of BP flakes that was not independently measured, the through-plane thermal conductivity of BP flakes, and the thermal conductance of BP/Si interface, see Ref. [11] for details. Jang *et al*. performed two TDTR measurements to derive the four unknowns simultaneously. However, due to the larger number of unknowns in Jang *et al.*'s analysis, higher uncertainty is expected. Particularly, in the Supporting Information of Ref. [11], Jang *et al.* reported that the derived thermal conductance of BP/Si interfaces varies by up to 50 % for their BP samples. This could be a source of errors in their derived $\Lambda_{TP}$ values.

To gain more insights on the $L_c$-dependent thermal conductivity in BP, we perform BTE calculations of heat conduction along the through-plane direction of BP by sandwiching BP between hot and cold reservoirs with a distance of $L_c$ apart, which provides additional boundary scattering mentioned above. Details of first-principles calculations and the frame work of phonon BTE are provided in Section S6 in the Supporting Information.

We note that the current first-principles calculations render the same thermal conductivity values of bulk BP as reported in Ref. [4]; the difference is that in this work, we extend the prediction to $L_c$-dependent thermal conductivity of BP by adding a boundary scattering term. The calculated through-plane thermal conductivity with an additional boundary scattering at $L_c$, as shown in Figure 5a, agrees well with both our frequency-dependent measurements and prior reported measurements on BP, for all three temperatures. As seen from Figure 5a, the through-plane thermal conductivity corresponding to $L_c$ = 10 nm, 100 nm and 1 µm is around 16 %, 53 % and 87 % of the bulk value, respectively, at 300 K. Our results thus provide a credible explanation to why Zhu *et al*.[4] obtained a $\Lambda_{TP}$ ~25 % smaller than the intrinsic value, i.e., they performed the measurements using $f$ = 9 MHz with the *2d* only ≈600 nm and thus heat transport by a substantial portion of nonequilibrium phonons were not measured in their experiments.

Since the $L_c$-dependent thermal conductivity comes from the long mean-free-path, in Figure 5b, we directly plot the thermal conductivity accumulation function[29] along TP axis at 80 K, 150 K and 300 K, calculated from our first-principles model, as a function of phonon mean-free-path $\ell$. (The accumulation functions along other axes are presented in the Supporting Information.) The accumulation functions are normalized by their corresponding bulk thermal conductivity.[4] The mean-free-path distribution at 300 K spans from several nanometers to several microns. The phonons with long mean-free-path phonons along TP direction should be related to low frequency acoustic phonons. For example, the mean-free-path of the LA phonon mode along TP direction with a frequency of 1 THz is around 1.6 µm, considering the sound velocity is around 4600 m/s from the calculated phonon dispersion and a phonon lifetime of 400 ps, as discussed below.

To understand the origins of the anisotropy in the thermal conductivity of BP, we first examine the phonon disperison of BP along the principal axes (ZZ, AC and TP) as shown in Figure S8 in the Supporting Information. The sound velocity (group velocity of low-frequency LA modes) is 8300, 4800, 4600 m/s along the three directions, showing anistropicity. The large difference of group velocity between ZZ and AC directions should be an important origin of the anisotropy of their thermal conductivity. We also notice that the difference of the sound velocity along AC and TP directions are small. However, from the phonon dispersion, the phonon branches other than LA branch along TP direction are rather flat compared with the two basal-plane directions. Therefore, the group velocity should be responsible for the anisotropy between TP direction and basal-plane directions to some extent.

Since thermal conductivity is also dependent on the phonon relaxation time (lifetime), we plot the phonon lifetime $\tau$ of the longitudinal (LA) and transverse (TA1 and TA2) phonons in BP along the principal axes (ZZ, AC and TP) in Figure 5c. Interestingly, we find that $\tau$ roughly scales with $\omega^2$ for LA and TA2 phonons but roughly scales with $\omega$ for TA1 phonons along all principal axis directions in the frequency range we calculated, but with different scattering strengths. We also observe that when both the vibrations and the propagation directions of phonons are within the basal planes (i.e., LA and TA2 phonons along ZZ and AC axes), the relaxation times of the phonon modes are rather isotropic in directions and are determined primarily by phonon frequency ω, see Figure 5c. The scattering of these in-plane vibration modes is relatively weaker. On the other hand, when either the vibrations or the propagation directions of phonons are out-of-plane (e.g., TA1 phonons, and phonons along the TP axis), phonons are much strongly scattered with clear anisotropy. The results thus suggest that the anisotropy of $\Lambda_{ZZ}$ and $\Lambda_{AC}$ in the basal planes is mainly due to anisotropy in the phonon dispersion, while the anisotropy of $\Lambda_{TP}$ and basal-plane thermal conductivity is due to both phonon dispersion and relaxation time.

In summary, we performed both frequency-dependent and beam offset TDTR measurements and the first-principles calculations on the thermal conductivity tensor of bulk BP in the temperature range of 80 – 300K. Our measurements and calculations provide consistent results. We derive the following important conclusions. 1) We observe a $T^{-1}$ dependence for the all three components of the thermal conductivity tensor and thus conclude that phonons are mainly scattered by the Umklapp processes in all crystallographic orientations in BP. 2) We obtain the intrinsic through-plane thermal conductivity of BP through frequency-dependent TDTR measurements and observe a considerable frequency dependence in the through-plane thermal conductivity measurements. Our measurements suggest that the phonon mean-free-paths are rather long in the through-plane direction. 3) From our first-principles calculations, we find that in BP, approximately, $\tau \propto \omega^2$ for LA and TA2 phonons, but $\tau \propto \omega$ for TA1 phonons in the frequency range we studied. Also $\tau$ is mostly isotropic in directions when both the vibrations and the propagation directions of phonons are in-plane (LA and TA2 phonons along ZZ and AC axes), but the scattering is strongly enhanced when either the vibrations or the propagation directions of phonons are out-of-plane (e.g., TA1 phonons, and phonons along the TP axis). We thus conclude that the anisotropy in the relaxation times only contributes to the anisotropy in the through-plane thermal conductivity, but not the anisotropy in the thermal conductivities along the basal planes. Our experimental and theoretical studies advance the fundamental understanding on heat transport in layered and highly anisotropic materials.

**Experimental Section**

*Growth of black phosphorus*: BP-2 and BP-3 were prepared using the following method. 20 mg Sn, 10 mg $SnI_4$ and 500 mg red phosphorus were weighed in a silica glass ampoule of 10 cm length, an inner diameter of 1.0 cm and a wall thickness of 0.25 cm. The ampoule was evacuated and placed horizontally in a dual zone split tube furnace, with the starting materials mixture located in the hot zone and the empty ampoule side in the colder zone. The reaction temperature was set to be 650 ˚C and 600 ˚C for hot zone and colder zone respectively and held for 1 h, and then cooled to 300 ˚C during 24 h for both zones.

*Polarized Raman spectroscopy*: Polarized Raman spectroscopy was conducted using a home-built micro-Raman setup. In our polarized Raman measurements, we used a p-polarized 532-nm continuous wave laser to excite the Raman spectra, and employed a linear polarizer before the spectrometer to collect only Raman-scattered light with polarization parallel to the polarization of the incident laser (i.e., the parallel-polarization configuration). The laser power was 3 mW and the spectrometer integration time was set to 2 s in all our measurements. More discussions on the Raman measurements can be found in Supporting Information.

In polarized Raman spectra, $B_{2g}$ mode shows a simple angular dependence, see Figure 1c, and is insensitive to laser wavelength and sample thickness;[11] thus, we use $B_{2g}$ peak to accurately determine the crystallographic orientations of BP. In previous studies,[11] the ZZ (or AC) axis was determined through continuous monitoring of the height of the $B_{2g}$ peak till the height is minimum. However, we find it difficult to pinpoint the minimum height. Instead, we conduct the polarized Raman measurements for every 15° and plot the integrated intensity of the $B_{2g}$ peaks as a function of rotation angle, see the red circles in Figure 1d. We then fit the measurements with $\sin^2(2(\theta+\alpha))$, where $\theta$ is the rotation angle and $\alpha$ is the correction angle (see Section 1 of the Supporting Information). The fitted $\alpha$ is usually small (in Fig 1d, $\alpha=0$) and can be used to accurately determine ZZ (or AC) axes. The uncertainty of our approach is ~2°, see the blue squares in Figure 1d. We then capture images of identifiable features on samples (e.g., the orientation of edges) by an in-situ bright-field microscope, to assist determination of the crystallographic orientations during thermal measurements. (Section 2 of the Supporting Information).

*TDTR and beam-offset TDTR*: In the beam-offset TDTR measurement, a pump beam with a small $w_0$ is modulated at a low frequency $f$, resulting in a wide in-plane heat spreading. To profile the temperature distribution generated by the pump beam, a probe beam is systematically offset in either the horizontal or the vertical directions while the out-of-phase signals at -100 ps are recorded, akin to measurements of laser spot sizes by autocorrelation,[13] see Figures 2a and 2c. (In most measurements, the horizontal and vertical directions correspond to the ZZ and AC axes of BP, respectively.) We then derived the in-plane thermal conductivity from the full-width-half-maximum (FWHM) of the beam-offset measurements by comparing the FWHM to calculations of an anisotropic thermal model,[13] see Section 5 of the Supporting Information for the iteration approach we applied to derive $\Lambda_{ZZ}$ and $\Lambda_{AC}$. As an internal consistency check, we compared two sets of measurements on BP-1, one performed when ZZ axis was horizontal and when it was vertical. We achieved an excellent agreement for both measurements, see Figure 2a and 2c. In addition, to ensure that our BP samples are properly passivated, we also performed another set of experiments 6 months after the initial sample preparation. We observed no changes in the beam-offset data, see Figure 2a and 2c, indicating that with the Al passivation, our BP samples did not deteriorate over time.

We also ensure that our approach to determine the crystallographic orientations by polarized Raman spectroscopy is accurate. We intentionally rotated the samples by as large as $\pm 30°$ and plot the derived thermal conductivity as a function of the angle relative to the ZZ axis, see Figure 2b and 2d. We find that the measurements form a symmetric ellipse with the major and minor axes of 83 W m$^{-1}$ K$^{-1}$ and 28 W m$^{-1}$ K$^{-1}$ at 300 K, as well as of 360 W m$^{-1}$ K$^{-1}$ and 130 W m$^{-1}$ K$^{-1}$ at 80 K, as expected from the rotation of the thermal conductivity tensor of BP.

More details on TDTR and beam-offset TDTR measurements and data analysis are presented in the Supporting Information.

**Acknowledgements**
This work is supported by the Singapore Ministry of Education Academic Research Fund Tier 2, under Award No. MOE2013-T2-2-147. R.Y. acknowledges the support from the US National Science Foundation (Grant No. 1512776) along with the Teets Family Endowed Doctoral Fellowship to X.G. Work conducted at NTU is financially supported by the Singapore National Research Foundation under NRF RF Award No. NRF-RF2013-08.

**Figure 1. Characterization of black phosphorus.** (a) Cystal structure of black phosphorus, in perspective, top and front views, respectively. Only two stacked layers are shown. (b) Optical image of our black phosphorus samples, the left was purchased from HQ-Graphene (BP-1) and the middle (BP-2) and the right (BP-3) were home-grown. BP-2 was coated with an Al thin film in the image. (c) Polarzied Raman spectra of BP-1 showing the out-of-plane $A_g^1$ mode and in-plane $B_{2g}$ and $A_g^2$ modes, acquired using the parallel-polarization configuration. The sample was rotated so that the angle between incident laser polarization and initial orientation of BP-1 ranges from 0° to 90°, as noted in the plot. Each plot is shifted by $3 \times 10^4$ Counts W$^{-1}$ s$^{-1}$ from previous measurement for clarity. (d) Angle dependence of the integrated intensity of the $B_{2g}$ peak shown in (c) (red circles), and when the measurements were repeated after we rotated the samples by 2° (blue squares). The black line is a fit of $\sin^2(2\theta)$ to the intensity of the $B_{2g}$ peak. (e) θ - 2θ XRD scan of BP-1.

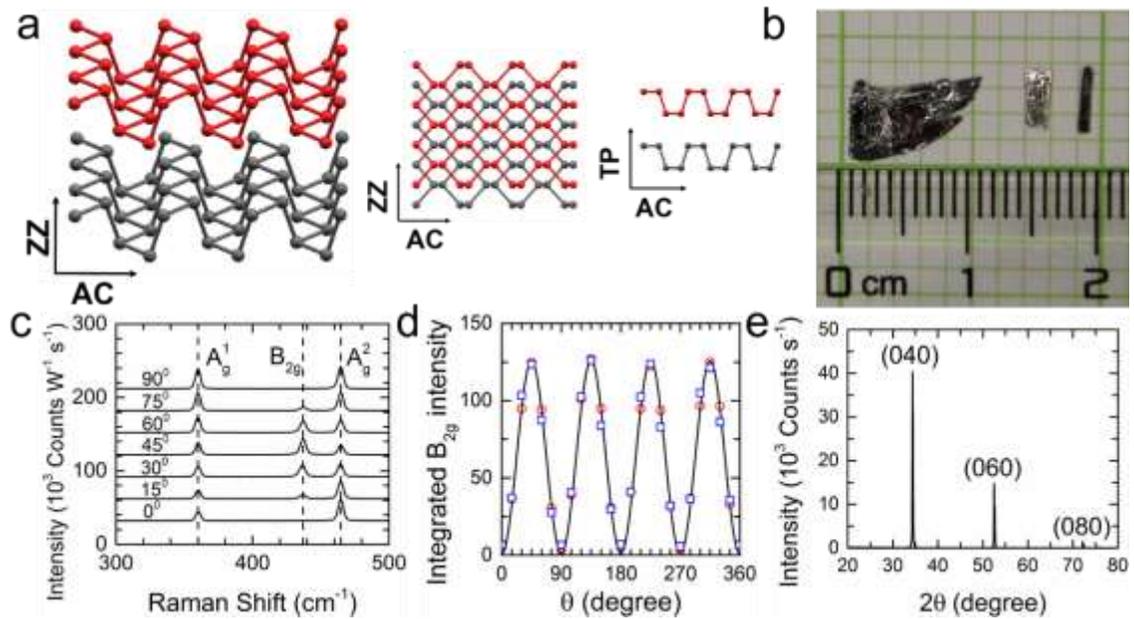

**Figure 2. Beam-offset TDTR measurements of BP-1 along the ZZ and AC axes**. (a) at 300 K using $w_0 = 2.5$ μm and $f = 0.5$ MHz, and (c) at 80 K using $w_0 = 5$ μm and $f = 0.5$ MHz. Three sets of measurements are plotted concurrently: one measured when the angle between ZZ axis and horizontal scan direction is 0° (red squares), one measured after the sample was rotated by 90° (blue triangles), and one after the sample was stored under ambient conditions for 6 months (black circles). The green lines are fits of an isotropic thermal model. (b, d) Thermal conductivity of black phosphorus at (b) 300 K and (d) 80 K derived from the beam-offset TDTR measurements on BP-1 (black circles), BP-2 (red triangles) and BP-3 (blue diamonds), as a function of the angle between the beam-offset scanning direction and the ZZ axis of the BP samples. The green lines are amplitudes of orthogonal components of the thermal conductivity tensor under rotation about the TP axis, with the major and minor axes of 83 W m⁻¹ K⁻¹ and 28 W m⁻¹ K⁻¹ at 300 K, and of 360 W m⁻¹ K⁻¹ and 130 W m⁻¹ K⁻¹ at 80 K.

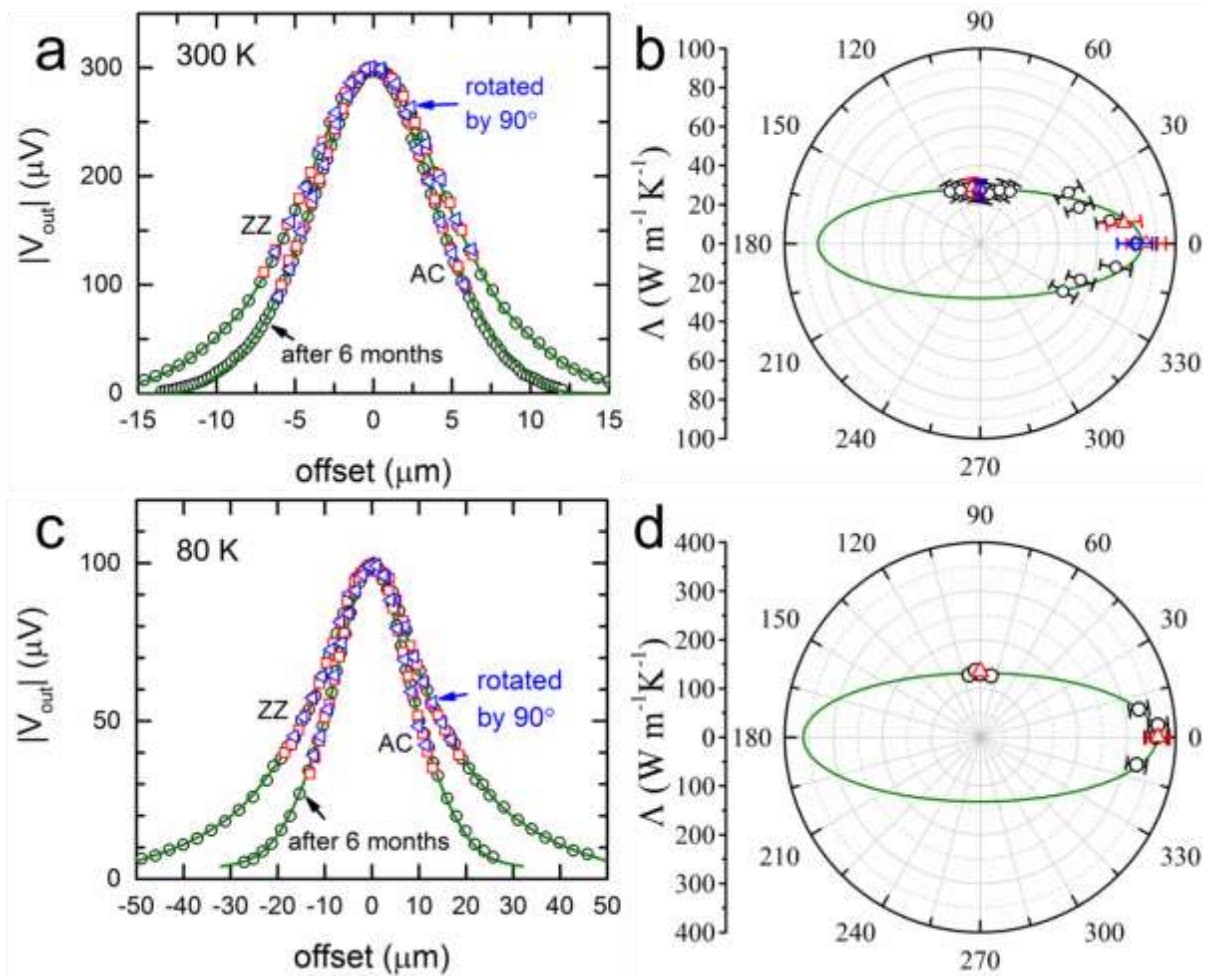

**Figure 3. Temeperature dependence of the anisotropic thermal conductivity tensor of black phosphorus**. Solid symbols represent data derived from measurements on BP-1 (black circles) and BP-2 (red triangles). All measurements were perfomed using $f = 0.5$ MHz. For measurements of $\Lambda_{TP}$, we used $w_0 = 25$ µm, while for measurements of $\Lambda_{ZZ}$ and $\Lambda_{AC}$, we used $w_0 = 5$ µm. For comparison, prior measurements of polycrystalline BP by Slack[30] (blue open cicles), of a 170-nm-thick BP nanoribbon by Lee *et al.*[6] in ZZ and AC directions (up and down open triangles) and of graphite by Nihira *et al.*[31] (orange open diamonds) are also plotted. Solid lines and dashed lines are the first-principles calculations of the thermal conductivity tensor of bulk BP and phosphorene, respectively, by Jain *et al.*[7] (green), Zhu *et al.* [4] (megenta) and Li *et al.*[8] (blue).

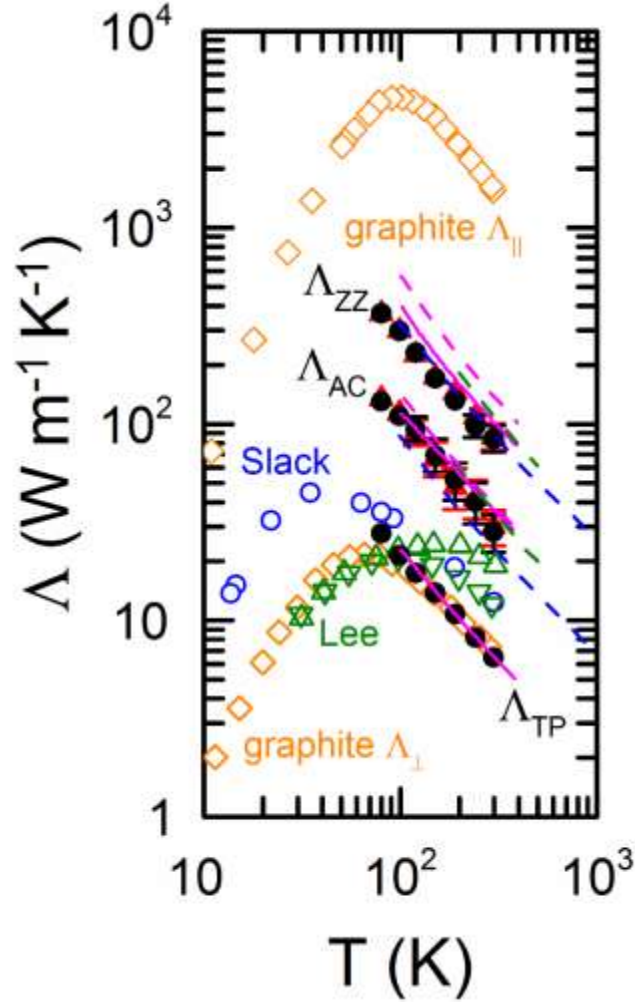

**Figure 4. Temeperature dependence of interfacial thermal conductance of Al/BP.** Our values (black circles) are nearly twice larger than that of TbFe/BP[4] (pink triangle) and NbV/AlO$_x$/BP[11] (red square). For comparasion, we also plot thermal conductance of Al/HOPG[24] (orange open diamonds), Al/single-layer-graphene(SLG)/Cu[32] (orange open circles) and Al/MoS$_2$[33] (blue open triangle).

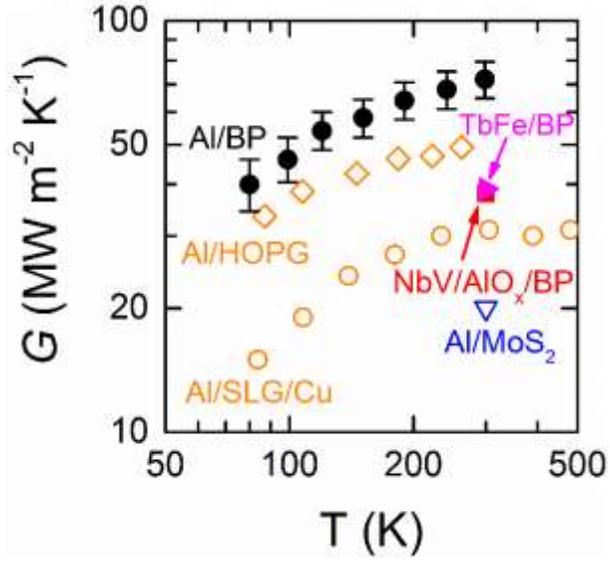

**Figure 5. Anisotropy in the mean-free-paths of phonons in BP.** (a) Frequency-dependent TDTR measurements of the through-plane thermal conductivity ($\Lambda_{TP}$) (circles) at 80 K (hollow), 150 K (half filled) and 300 K (solid), compared to measurements on BP flakes with thickness of 138 - 552 nm by Jang et al.[11] (ssquares), measurements of BP at 9.1 MHz by Zhu et al.[4] (triangles), and our first-principles calculations on $L_C$-dependent $\Lambda_{TP}$ at 80 K (dash-dot line), 150 K (dashed line) and 300 K (solid line). All measurements and calculations are normalized by the calculated $\Lambda_{TP}$ of bulk BP of 6.5 W m$^{-1}$ K$^{-1}$ (300 K), 14 W m$^{-1}$ K$^{-1}$ (150 K) and 30 W m$^{-1}$ K$^{-1}$ (80 K) respectively. All measurements are plotted as a function of a charateristic length $L_C = 2d$ or flake thickness $h$, whichever smaller, as defined in main text. (b) Thermal conductivity accumulation along through-plane (TP) directions at 300 K (solid line), 150 K (dashed line) and 80 K (dash-dot line) from our first-principles calculations. The accumulated thermal conductivity calculated from the first-principles calculations is normalized by the respective bulk thermal conductivity. (c) The relaxation times ($\tau$) of longitudial (LA) and transverse (TA1 and TA2) phonon modes along the high-symmetry axes of BP, as a function of angular frequency ($\omega$) of phonons.

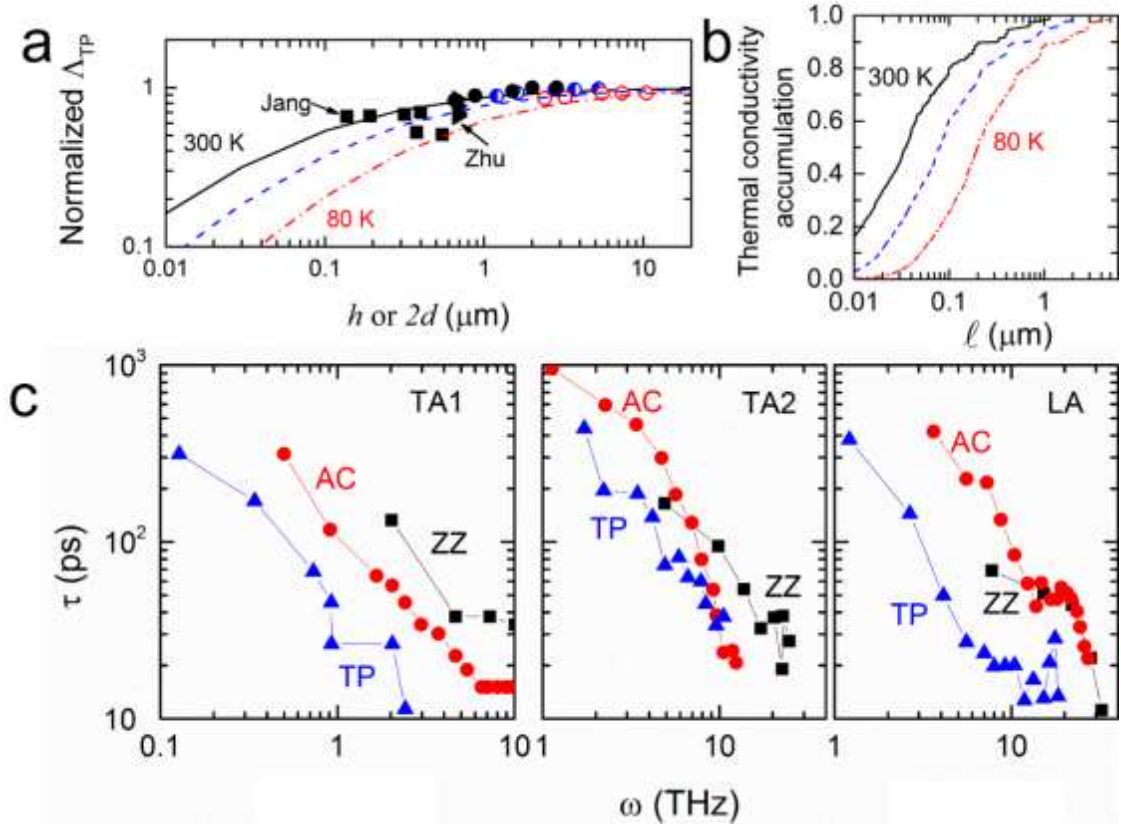

**Table 1. Summary of all the measurements at 300 K.** The unit is W m$^{-1}$ K$^{-1}$.

|      | $w_0$=2.5 μm, $f$=0.5MHz | | $w_0$=5 μm, $f$=0.5MHz | | $w_0$=25 μm, $f$=0.5MHz |
|------|:---:|:---:|:---:|:---:|:---:|
|      | $\Lambda_{ZZ}$ | $\Lambda_{AC}$ | $\Lambda_{ZZ}$ | $\Lambda_{AC}$ | $\Lambda_{TP}$ |
| **BP-1** | 80.4 ± 10 | 26.4 ± 5 | 84 ± 11 | 28 ± 6 | 6.5 ± 0.8 |
| **BP-2** | 85 ± 10 | 28 ± 5 | 86 ± 11 | 30 ± 6 | |
| **BP-3** | 79.2 ± 10 | 27.6 ± 5 | | | |

Supporting Information

Temperature dependence of anisotropic thermal conductivity tensor of bulk black phosphorus

*Bo Sun, Xiaokun Gu, Qingsheng Zeng, Xi Huang, Yuexiang Yan, Zheng Liu, Ronggui Yang and Yee Kan Koh\**

# Table of Contents:

## S1. Polarized Raman measurements

*Angular dependent polarized Raman peaks of BP.*

The Raman tensors of $A_g$ modes and $B_{2g}$ mode are written as

$$R_{A_g} = \begin{pmatrix} a & 0 & 0 \\ 0 & b & 0 \\ 0 & 0 & c \end{pmatrix}$$

$$R_{B_{2g}} = \begin{pmatrix} 0 & 0 & f \\ 0 & 0 & 0 \\ f & 0 & 0 \end{pmatrix}$$

According to previous publications,[1, 2] the polarized Raman intensities of $A_g$ modes and $B_{2g}$ mode under parallel-polarization configuration are

$$I_{A_g} \propto \left(|a|\sin^2\theta + |c|\cos\phi_{ca}\cos^2\theta\right)^2 + |c|^2 \sin^2\phi_{ca} \cos^4\theta \tag{s1}$$

$$I_{B_{2g}} \propto (|f|\sin 2\theta)^2 \tag{s2}$$

where θ is the angle between ZZ axis and the polarization of incident laser beam, $\phi_{ca}$ the phase difference between c and a, as both are complex numbers with phases.

We learn that the angular dependent intensities of $A_g$ modes are quite complex and are sensitive to both BP thickness and laser-wavelength, while $B_{2g}$ mode show a simple $\sin^2(2\theta)$ dependence. Detailed discussions of the angular dependence behavior of $A_g$ modes and $B_{2g}$ mode can be found in SI Reference [2]. Here we will not go to details as our purpose is to identify the crystal orientation using polarized Raman spectra.

The angular dependent intensity of $B_{2g}$ peaks can be fit by $\sin^2(2\theta)$. We take note that θ is the angle between ZZ axis and incident laser beam polarization, not the rotation angle from when the height of $B_{2g}$ peak is minimum. There is normally a small correction angle α between ZZ axis and the initial orientation when the height of $B_{2g}$ peak is minimum, since the minimum is determined subjectively and with large uncertainty. We then fit the intensities of $B_{2g}$ peaks by $\sin^2(2(\theta+\alpha))$, and the resulting α could help us find the real ZZ axis with rather small uncertainty.

***Homogeneity of BP-1.*** Since BP-1 is quite large, we checked its homogeneity by conducting polarized Raman measurements on 4 randomly chosen points at corners of BP-1 sample. The Raman spectra are shown in Figure S1. The intensities of $B_{2g}$ peaks are zero and the ratios of integrated $A_g^2$ and $A_g^1$ peaks are the same, indicating that the crystallographic orientations are the same across the sample.

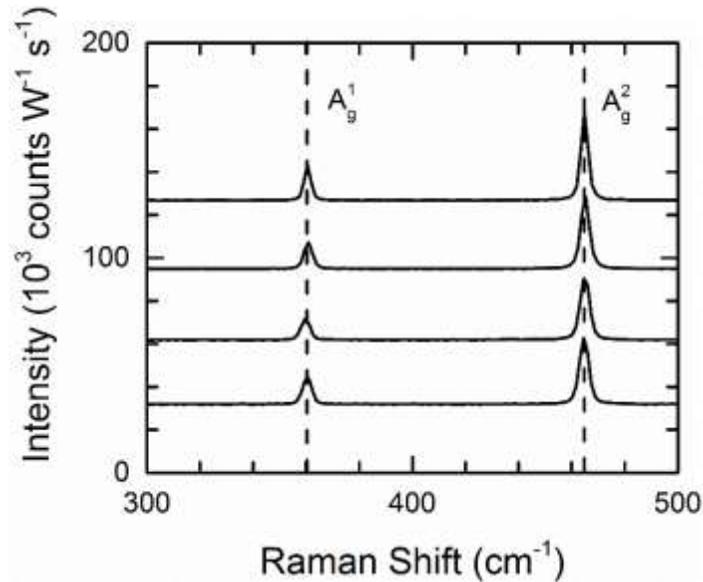

**Figure S1.** Raman spectra of randomly chosen points of BP-1 sample. The points are located at the four corners of BP-1. Each measurement shifts up $3\times10^4$ counts/W/s from previous measurement for clarity.

## S2. Transfer samples to TDTR stage

After ZZ axis is identified using the polarized Raman spectroscopy, we have to move the samples to TDTR stage for thermal measurement. In Raman setup, we have determined that the ZZ axis is parallel to the laser polarization direction, i.e. horizontal. In TDTR setup, normally we set the horizontal (vertical) scan along ZZ (AC) direction. (By saying horizontal scan, we mean that pump and probe beams offset horizontally from one side to the other side.) So the samples should be moved from Raman stage to TDTR stage with the crystal orientation maintained.

We make use of the sample edges to help us identify the sample orientations. First, we take an image using bright-field microscope in our Raman setup after the crystal orientations are identified (Figure S2a). One edge is chosen and the angle between it and the horizontal reference line is 36°. Second, we move the sample to TDTR stage, and take another image. The same edge will be found and the angle between it and the horizontal reference line should maintain the same (Figure S2b).

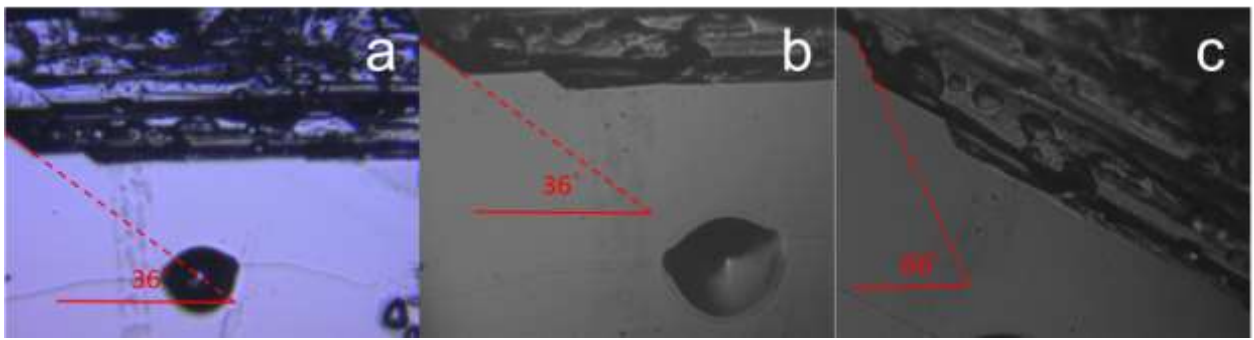

**Figure S2.** Bright field microscope image of sample BP-1 taken at Raman setup (a) and after transferred to TDTR setup (b). The same edge is used and the angle between it and the horizontal reference line (red solid lines) are maintained the same. We also rotate the sample by up to ±30° to do TDTR measurements. An example is shown in (c) when the sample is rotated by 30°.

## S3. TDTR and beam-offset TDTR measurements

Details of our implementation, data analysis and signal processing of TDTR are presented in SI Reference [3]. For beam-offset TDTR, the method has been well established and described in SI Reference [4].

In TDTR, laser pulses are split into a pump and a probe beams. The pump beam, modulated at frequency $f$, heats the sample periodically. With the periodic heating, a temperature oscillation at modulation frequency $f$ is induced within a distance $d$ from sample surface. ( $d = \sqrt{\Lambda/\pi C f}$ is called the thermal penetration depth, where $\Lambda$ is the thermal conductivity and $C$ is the volumetric heat capacity.) We then monitor the temperature oscillation at the sample surface by the time-delayed probe beam. $\Lambda_{TP}$ is derived by comparing the measured temperature oscillation to calculations of an anisotropic thermal model with derived $\Lambda_{ZZ}$ and $\Lambda_{AC}$.[4] As mentioned in the main text, we use a spot size of $w_0$= 25 μm and modulation frequency of pump beam $f$ of 0.5 MHz. We also use $f$ of 1 MHz, 2 MHz, 5 MHz, 10 MHz for frequency dependent measurement of $\Lambda_{TP}$. The pump beam power was kept at 100 mW and probe beam power was ~40 mW, which lead to a ~4 K and ~1 K temperature rise at room temperature and ~79 K, respectively. We use our newly developed pump leak correction approach[3] to eliminate the artificial signals reflected from the cryostat window.

We measure $\Lambda_{ZZ}$ and $\Lambda_{AC}$ of BP by the beam-offset TDTR method.[4] Unlike previous measurements using NbV[5] and TbFe[6] as transducer, we use Al as transducer for measurements at 80 K – 300 K due to its heat capacity is well reported within the temperature range. We use a power of ~20 mW for pump beam and ~6 mW for probe beam, which will result in similar steady-state temperature rises to those in TDTR measurements.

# S4. Measurement sensitivity and uncertainty analysis in TDTR and beam-offset TDTR

For conventional TDTR, the ratio $R=-V_{in}/V_{out}$ is used to extract the thermal conductivity, thus, the sensitivity parameter $S_\alpha$ is defined as[7]

$$S_\alpha = \frac{\partial \ln R}{\partial \ln \alpha}$$

where α is the parameter used in the thermal model for TDTR, such as the thermal conductivity of substrate Λ, the thickness of Al, or the heat capacity of substrate.

For beam-offset TDTR, the FWHM of out-of-phase signal $V_{out}$ is used to extract the in-plane thermal conductivities.[4] In such case, the sensitivity parameter $S_\alpha^{FWHM}$ is defined as

$$S_\alpha^{FWHM} = \frac{\partial \ln(\text{FWHM})}{\partial \ln \alpha}$$

where α is defined the same as above.

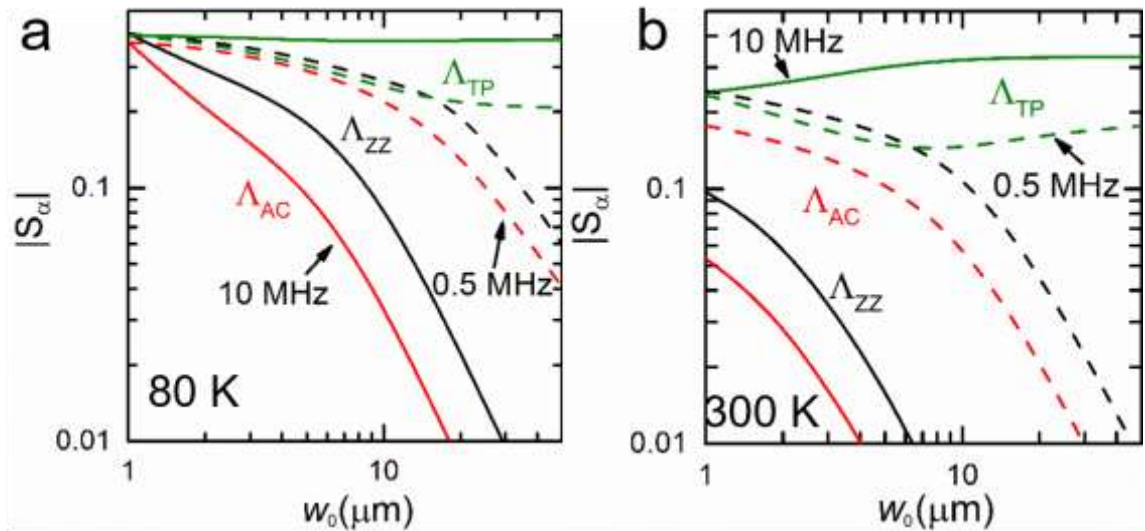

**Figure S3**. Sensitivity of TDTR signals to components of the thermal conductivity tensor along TP, ZZ and AC axes at 80 K (a) and at 300 K (b). We assume laser $1/e^2$ radii $w_0$ of 1-50 μm and modulation frequencies $f$ of 10 MHz (solid lines) and 0.5 MHz (dased lines), respectively, in both figures.

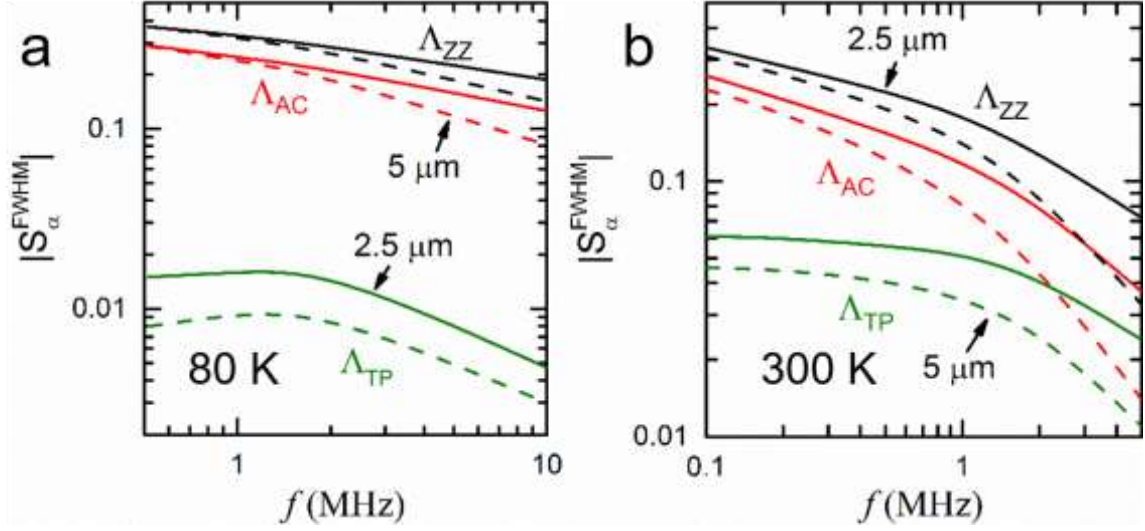

**Figure S4**. Sensitivity of FWHM in the beam-offset TDTR measurements, to components of the thermal conductivity tensor along TP, ZZ and AC axes, at 80 K (a) and at 300 K (b). we assume $f = 0.1\text{-}10$ MHz and $w_0 = 2.5$ μm (solid lines) and 5 μm (dased lines), respectively. At 300K, the sensitivities of FWHM to $\Lambda_{ZZ}$ and $\Lambda_{AC}$ are 0.227 and 0.157, repectively, for $f=0.5$MHz and $w_0=2.5$ μm. These values are larger than those in previous measurements by Jang et al.[5] (0.15 and 0.11) and by Zhu et al.[6] (0.21 and 0.14) using $f=1.6$ MHz and $w_0 \sim 2.5$ μm, since our $f$ is much lower than theirs, altough the Al transduer we used has a higher thermal conductantity than NbV used by Jang et al.[5] and TbFe used by Zhu et al.[6].

After calulations of measurement sensitivities to all parameters are finished, we then evaluate the measurement uncertainty of $\Lambda_{TP}$, $\Lambda_{ZZ}$ and $\Lambda_{AC}$ using the following equation

$$\left(\frac{\Delta\Lambda_{TP}}{\Lambda_{TP}}\right)^2 = \sum\left(\frac{S_\alpha}{S_{\Lambda_{TP}}}\frac{\Delta\alpha}{\alpha}\right)^2 + \left(\frac{S_\phi}{S_{\Lambda_{TP}}}\delta\phi\right)$$

$$\left(\frac{\Delta\Lambda_{ZZ}}{\Lambda_{ZZ}}\right)^2 \propto \left(\frac{\Delta FWHM_{ZZ}}{FWHM_{ZZ}}\right)^2 = \sum\left(\frac{S_\alpha^{FWHM}}{S_{\Lambda_{ZZ}}^{FWHM}}\frac{\Delta\alpha}{\alpha}\right)^2 \quad \text{(s3)}$$

$$\left(\frac{\Delta\Lambda_{AC}}{\Lambda_{AC}}\right)^2 \propto \left(\frac{\Delta FWHM_{AC}}{FWHM_{AC}}\right)^2 = \sum\left(\frac{S_\alpha^{FWHM}}{S_{\Lambda_{AC}}^{FWHM}}\frac{\Delta\alpha}{\alpha}\right)^2$$

where $\Delta\alpha/\alpha$ is the uncertainty of parameter α, $\delta\phi$ is the phase uncertainty. Please note that the uncertainty of phase is not considered in beam-offset TDTR measruements, since the the out-of-phase signal at -100ps is insentivite to phase, which result in a much lower measurement sensitivity of FWHM to phase comparing with TDTR measurements.

For beam-offset TDTR, we first calculate the uncertainty of FWHM. The uncertainty of FWHM will be translated to uncertainty of $\Lambda_{ZZ}$ and $\Lambda_{AC}$ by FWHM – $\Lambda_{ZZ}$ and FWHM – $\Lambda_{AC}$ plots. An example of such plot is shown in Figure S6 in next section.

## S5. Data analysis for beam-offset TDTR measurements

The beam-offset TDTR will lead to a measurement as in Figure 2a, where FWHM of $V_{out}$ is used to calculate $\Lambda_{ZZ}$ and $\Lambda_{AC}$. To extract $\Lambda_{ZZ}$ and $\Lambda_{AC}$, we first fit the measurements using an isotropic thermal model,[4, 8] *i.e.* assuming $\Lambda_{ZZ} = \Lambda_{AC}$, which is shown as green lines in Figure 2a. Such fit will give us the initial values of $\Lambda_{ZZ}$ and $\Lambda_{AC}$ ($\Lambda_{ZZ, 1} = 87.5$ W m$^{-1}$ K$^{-1}$, $\Lambda_{AC, 1} = 24.2$ W m$^{-1}$ K$^{-1}$). This is the first iteration.

Although the sensitivity to $\Lambda_{AC}$ ($\Lambda_{ZZ}$) of the FWHM along ZZ (AC) direction in beam-offset measurement is rather small (for example, the sensitivity to $\Lambda_{ZZ}$ is 0.227 while is 0.019 to $\Lambda_{AC}$ when pump and probe beams offset along ZZ direction for the case of Figure 2a), the anisotropy of in-plane thermal conductivity should not be neglected. Thus, we calculate the FWHM using the anisotropic thermal model[4] with $\Lambda_{ZZ,1}$ and $\Lambda_{AC,1}$ as the input, the results of second iteration are shown in Figure S5.

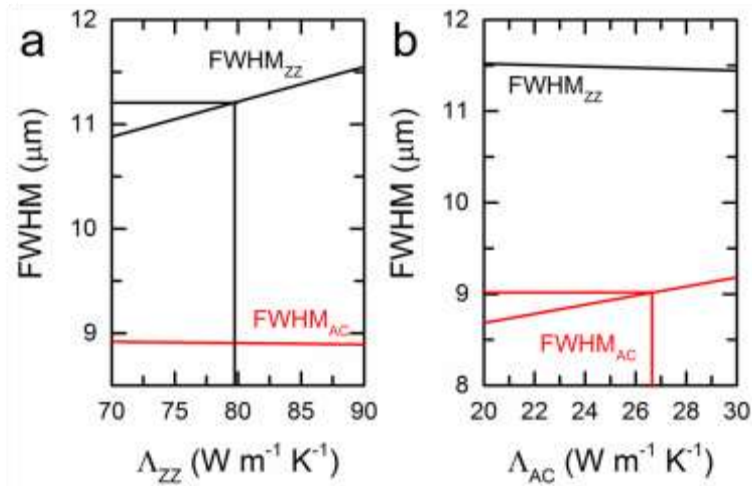

**Figure S5**. The 2$^{nd}$ iteration: Calculated FWHMs when pump and probe beams offset along ZZ (FWHM$_{ZZ}$) and AC (FWHM$_{AC}$) directions. The calculation is based on the anisotropic thermal model with $\Lambda_{ZZ, 1} = 87.5$ W m$^{-1}$ K$^{-1}$ and $\Lambda_{AC, 1} = 24.2$ W m$^{-1}$ K$^{-1}$ as input. The experimental values (FWHM$_{ZZ}$ =11.20 µm and FWHM$_{AC}$ =9.03 µm) are used to extract $\Lambda_{ZZ, 2}$ and $\Lambda_{AC, 2}$.

By comparing the FWHM we measured with those calculated, we then get the second set of $\Lambda_{ZZ}$ and $\Lambda_{AC}$ ($\Lambda_{ZZ, 2} = 79.7$ W m$^{-1}$ K$^{-1}$, $\Lambda_{AC, 2} = 26.6$ W m$^{-1}$ K$^{-1}$), which is ~10% larger or smaller than the initial values we get from the isotropic assumption.

Then, for the 3$^{rd}$ iteration, we put the second data set $\Lambda_{ZZ, 2}$ and $\Lambda_{AC, 2}$ as the input in the anisotropic thermal model, the calculated FWHM is shown in Figure S6. The third set of $\Lambda_{ZZ}$ and $\Lambda_{AC}$ will be extracted, and the values ($\Lambda_{ZZ, 3}$ = 80.4 W m$^{-1}$ K$^{-1}$, $\Lambda_{AC, 3}$ = 26.4 W m$^{-1}$ K$^{-1}$) have a difference of <1% with $\Lambda_{ZZ, 2}$ and $\Lambda_{AC, 2}$.

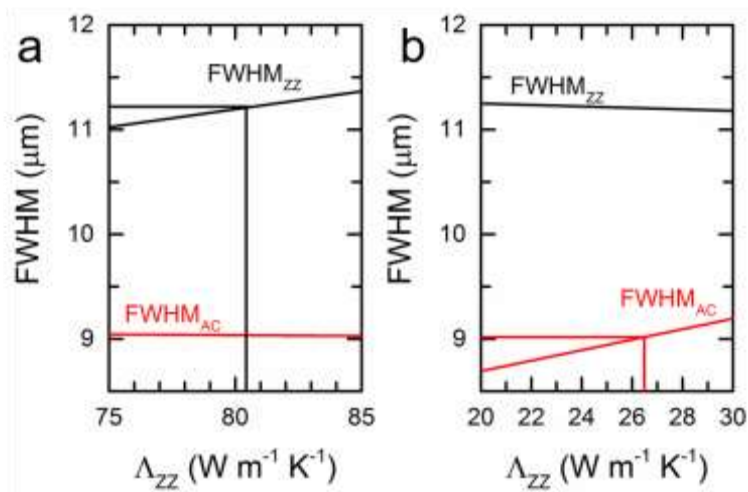

**Figure S6**. The 3$^{rd}$ iteration: Calculated FWHMs when pump and probe beams offset along ZZ (FWHM$_{ZZ}$) and AC (FWHM$_{AC}$) directions. The calculation is based on the anisotropic thermal model with $\Lambda_{ZZ, 2}$ = 79.7 W m$^{-1}$ K$^{-1}$ and $\Lambda_{AC, 2}$ = 26.6 W m$^{-1}$ K$^{-1}$ as input. The experimental values (FWHM$_{ZZ}$ =11.20 µm and FWHM$_{AC}$ =9.03 µm) are used to extract $\Lambda_{ZZ}$ and $\Lambda_{AC}$.

Since $\Lambda_{ZZ, 3}$ and $\Lambda_{AC, 3}$ are already very close to $\Lambda_{ZZ, 2}$ and $\Lambda_{AC, 2}$, another iteration for calculation of FWHM using the anisotropic model with $\Lambda_{ZZ, 3}$ and $\Lambda_{AC, 3}$ as input are unnecessary. Nevertheless, we still calculate them and find that it gives the same thermal conductivities as $\Lambda_{ZZ, 3}$ and $\Lambda_{AC, 3}$.

All together, we need at least 2 iterations to get the final results of $\Lambda_{ZZ}$ and $\Lambda_{AC}$: the first one calculated using isotropic thermal model with the following 1 or 2 iterations using anisotropic thermal model. These steps apply to measurements at all temperatures since the aspect ratios of $\Lambda_{ZZ}$ and $\Lambda_{AC}$ are similar. Since the calculation using anisotropic thermal model is compute-intensive (calculating one FWHM needs several hours to days depending on the heat penetration depth. For example, each point calculated above need 2 hours on dual hexa-core Intel X5650 2.66GHz processors using MATLAB parallel computing toolbox), 2 iterations will be good enough to get $\Lambda_{ZZ}$ and $\Lambda_{AC}$ with computing uncertainties <1%.

# S6. First-principles calculation for phonon transport

## S6.1 Thermal conductivity from Peierls-Boltzmann transport equation

We performed first-principles-based Peierls-Boltzmann transport equation (PBTE) calculations to predict the thermal conductivity of both bulk and monolayer black phosphorus.

The thermal conductivity of a crystal material along α direction is expressed as

$$\Lambda_{\alpha\alpha} = \frac{1}{N_0 \Omega} \sum_{qs} \hbar \omega_{qs} v_{qs}^{\alpha} n_{qs}^0 (n_{qs}^0 + 1) F_{qs}^{\alpha}, \qquad (s4)$$

where $\Omega$ is the volume of primitive unit cell, $\hbar$ is the Planck constant, $\omega_{qs}$ and $v_{qs}$ are the frequency and group velocity of phonon mode $qs$. $N_0$ is the number of q points. The phonon frequency $\omega_{qs}$, group velocity $v_{qs}$ ($=\nabla\omega_{qs}$) and equilibrium phonon population $n_{qs}^0$ are obtained from the phonon dispersion of the crystal, which is related to the second-order harmonic force constants of the crystal, $\phi$. $F$ is the deviation function describing how the non-equilibrium phonon population $n$ from $n^0$ through $n_{qs} = n_{qs}^0 + n_{qs}^0(n_{qs}^0 + 1)F_{qs}^{\alpha}$.

$F$ is solved from PBTE, which is expressed as[9, 10]:

$$v_{qs}^{\alpha} \frac{\partial n_{qs}^0}{\partial T} = \sum_{q's',q''s''} \left[ W_{qs,q's'}^{q''s''} \left( F_{q''s''}^{\alpha} - F_{q's'}^{\alpha} - F_{qs}^{\alpha} \right) + \frac{1}{2} W_{qs}^{q's',q''s''} \left( F_{q''s''}^{\alpha} + F_{q's'}^{\alpha} - F_{qs}^{\alpha} \right) \right] - \frac{n_{qs}^0(n_{qs}^0+1)F_{qs}^{\alpha}}{L_c/2|v_{qs}^{\alpha}|}, \qquad (s5)$$

where $W_{qs,q's'}^{q''s''}$ and $W_{qs}^{q's',q''s''}$ are the equilibrium transition probabilities for three-phonon annihilation and decay scattering processes, respectively. $W_{qs,q's'}^{q''s''}$ and $W_{qs}^{q's',q''s''}$ are determined by the third-order anharmonic force constants, and their expressions can be found in Ref. [10]. The last term represents the additional boundary scattering due to frequency modulation, as discussed in main text.

With phonon dispersion and deviation function $F$, the thermal conductivity is computed through Eq. (s4). In the calculation, we sampled the first Brillouin zone using a $N\times N\times N$ q-mesh.

**S6.2 Extracting interatomic force constants from first-principles calculations**

We perform first-principles calculations to extract the second-order harmonic and third-order anharmonic force constants, which are used to determined phonon dispersion and the phonon-phonon scattering rates for computing the thermal conductivity.

Our first-principles calculations were carried out using the Vienna *ab initio* Simulation Package (VASP).[11] The projector augmented wave pseudopotential with PBE functional[12] is employed. We note that for bulk black phosphorus obtaining accurate treatment of van der Waals (vdW) interaction in DFT calculations is still a hot research topic. But there have been many approaches proposed to deal with wdW interaction, and a few of them have been widely used. In this work, we took into account the van der Waals interaction by using the dispersion correction (DFT-D) proposed by Grimme[13]. This method had been tested for mechanical properties of BP and was regarded as the best among many other treatments[14]. We used periodic boundary conditions throughout the study. The kinetic-energy cut-off for the plane-wave basis set is set to be 500 eV and a 12×12×6 Monkhorst-Pack mesh is used to sample the reciprocal space of bulk (monolayer) black phosphorus. When we further refine these two parameters, the energy change is smaller than 1 meV/atom. The crystal structure (See Fig. S7), including the lattice constants and atom coordinates, is relaxed through the conjugate gradient algorithm until the stress within the material is zero and the atomic forces are smaller than $1\times10^{-5}$ eV/Å. The optimized lattice constants, which are denoted as *a*, *b* and *c* for the zigzag, through-plane and armchair directions, respectively, are summarized in Table S1.

**Table S1**. Lattice constants for both bulk and monolayer black phosphorus

|  | Method | $a$ (Å) | $b$ (Å) | $c$ (Å) |
|---|---|---|---|---|
|  | DFT-D (this work) | 3.321 | 10.477 | 4.427 |
| Bulk | DFT-D (Ref. [14]) | 3.30 | 10.43 | 4.40 |
|  | Exp. (Ref. [15]) | 3.3133 | 10.473 | 4.374 |

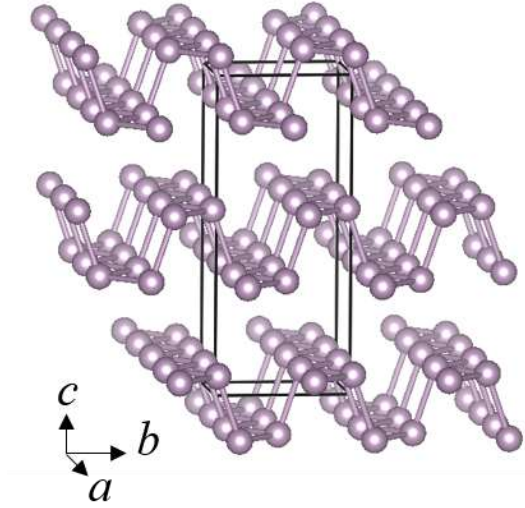

**Figure S7**. Crystal structure of bulk black phosphorus.

With the obtained equilibrium crystal structures, we extracted the interatomic force constants using the supercell-based method in the first-principles calculations.[16] The force component in $\alpha$ direction on each atom in a supercell is expressed as[16]

$$F_{R\tau}^{\alpha} = -\sum_{R'\tau'\beta} \phi_{R\tau,R'\tau'}^{\alpha\beta} u_{R'\tau'}^{\beta} - \frac{1}{2}\sum_{R'\tau'\beta}\sum_{R''\tau''\gamma} \psi_{R\tau,R'\tau',R''\tau''}^{\alpha\beta\gamma} u_{R'\tau'}^{\beta} u_{R''\tau''}^{\gamma} + \cdots \qquad (s6)$$

where $u$ is displacement of an atom away from its equilibrium position. When calculating the second-order harmonic force constants, we displaced one atom in a supercell by a small displacement $\Delta u = 0.02$ Å away from its equilibrium position along $\pm x$, $\pm y$ and $\pm z$ directions, and then record the forces of all atoms in the supercell. With the recorded forces, we extract the second-order harmonic force constants by fitting the displacement-force data set according to Eq. (s6).

The cutoff of the harmonic interactions is chosen to be 3.0 $a$ (~ 10 Å). In order to take the interlayer interaction into account, more than one layers should be included in the supercell, leading to the computational challenges due to the large amount of atoms in the simulation. To avoid employing big supercells to obtain the displacement-force data corresponding to the long-range interlayer interaction, we use two kinds of small supercells with different dimensions (6×6×1 and 4×4×2 conventional unit cells) to generate two sets of displacement-force data sets, which are fitted simultaneously to extract the second-order harmonic force constants. With the extracted harmonic force constants, the phonon dispersion of black phosphorus is calculated, which is presented in Fig.S8. The calculated phonon dispersion of bulk black phosphorus agrees reasonably well with the experimental data from the inelastic neutron scattering measurements[17].

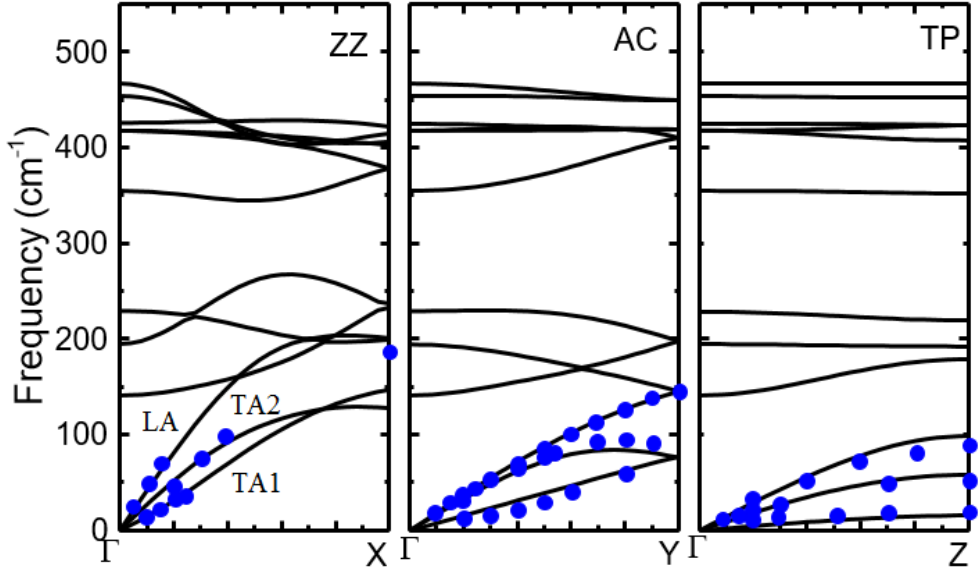

**Figure S8. Phonon dispersion of bulk BP.** Black lines are the calculated phonon dispersion using the second-order harmonic force constants extracted from first-principles calculations. Blue dots are the experimental results from inelastic neutron scattering measurement [17].

Similarly, to extract the third-order anharmonic force constants for bulk black phosphorus, we displaced two atoms in a 4×3×2 conventional unit cell with a distance of 0.02 Å along different directions simultaneously. With the recorded force information on all atoms in the supercell, the third-order force constants was then calculated using the finite-difference scheme[18]:

$$\psi^{\alpha\beta\gamma}_{R\tau,R'\tau',R''\tau''} = \frac{1}{4\Delta u^2}\left[-F^{\alpha}_{R\tau}\left(u^{\beta}_{R'\tau'}=\Delta u, u^{\gamma}_{R''\tau''}=\Delta u\right) + F^{\alpha}_{R\tau}\left(u^{\beta}_{R'\tau'}=\Delta u, u^{\gamma}_{R''\tau''}=-\Delta u\right) + F^{\alpha}_{R\tau}\left(u^{\beta}_{R'\tau'}=-\Delta u, u^{\gamma}_{R''\tau''}=\Delta u\right) - F^{\alpha}_{R\tau}\left(u^{\beta}_{R'\tau'}=-\Delta u, u^{\gamma}_{R''\tau''}=-\Delta u\right)\right]. \quad (s7)$$

As the calculated thermal conductivity is highly sensitive to the cutoff used for the calculation of the third-order force constants, we carefully tested the choice of cutoff. Fig. S9 shows the calculate thermal conductivity of bulk black phosphorus with different cutoff using a 13×13×13 q-point phonon mesh on the first Brillouin zone. It is clearly seen that a too small cutoff could substantially overestimate the thermal conductivity along all directions. When the cutoff is larger than 5.55 Å, the thermal conductivity variation is smaller than 10%. Therefore, we chose 5.55 Å as the cutoff for anharmonic third-order force constants for further calculations.

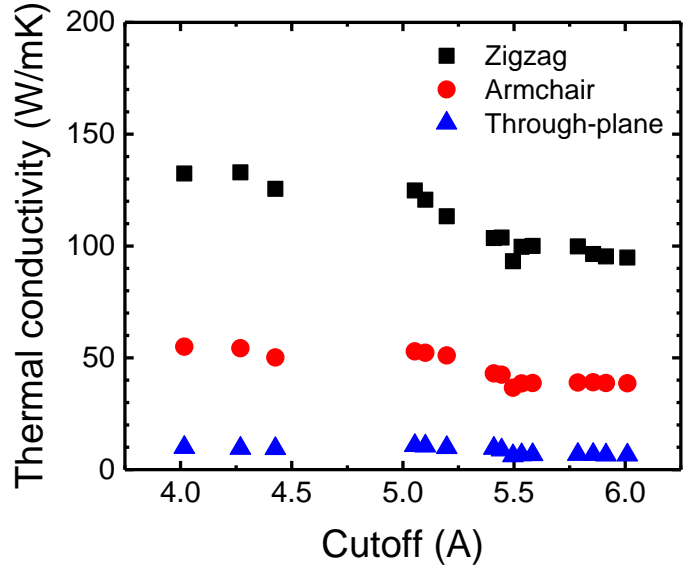

**Figure S9**. The calculated thermal conductivity of bulk black phosphorus as a function of the cutoff used for the calculation of third-order anharmonic force constants.

With the interatomic force constants, the thermal conductivity of BP can be calculated using the Boltzmann transport equation method. To account the contributions from different phonon modes, the first Brillouin zone is sampled using finite number of q points. To ensure the q mesh employed lead to converged thermal conductivity, we calculate the thermal conductivity of MoS$_2$, as an example, using q-meshes with $N*N*N$ points, as shown in Fig. S10. When the q mesh is denser than 15*15*15, the thermal conductivity is almost constant. Therefore, we used $N = 61$ and

$N = 19$ in the remaining thermal conductivity calculations for monolayer and bulk black phosphorus, respectively.

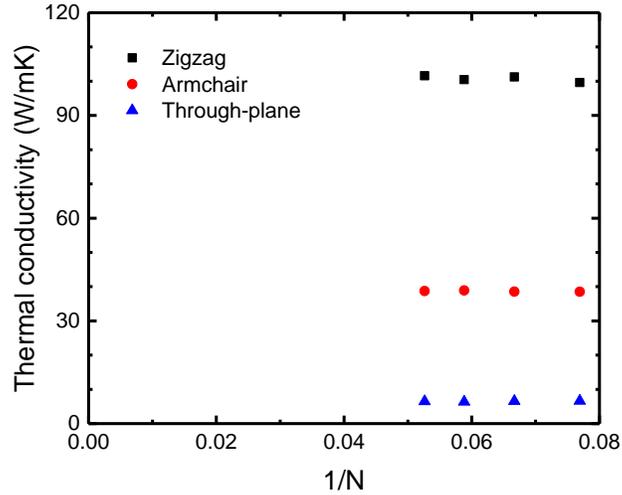

**Figure S10**. The calculated thermal conductivity of black phosphorus as a function of the ($N \times N \times N$) q-point mesh. We test the meshes with 13×13×13, 15×15×15, 17×17×17, 19×19×19 points for bulk black phosphorus.

We plot the thermal conductivity accumulation function along three principal axes, calculated from our first-principles model, as a function of phonon mean-free-path $\ell$. The mean-free-path distributions along all directions span a similar range, although thermal conductivities along these directions vary by more than an order of magnitude.

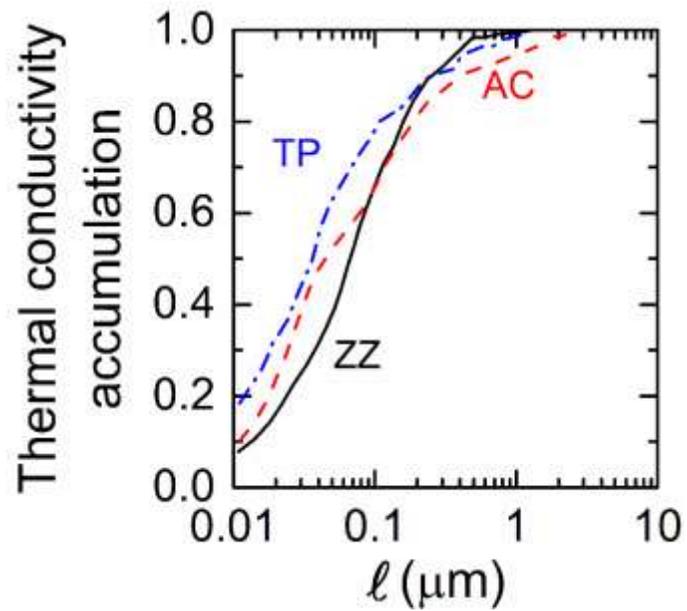

**Figure S11**. Thermal conductivity accumulation along zigzag (ZZ), armchair (AC) and through-plane (TP) directions from our first-principles calculations at 300 K. The accumulated thermal conductivity is normalized by the respective bulk thermal conductivity.

## S7. Temperature dependent heat capacity of BP

In the data analysis of TDTR measurements, we used the same source for the volumetric heat capacity of BP,[19] as what were previously used by Zhu *et al*.[6] and Jang *et al*.[5] The volumetric heat capacity from Ref. [19] is plotted in Fig. S12. At room temperature, the volumetric heat capacity C is 1.87 J cm$^{-3}$ K$^{-1}$.

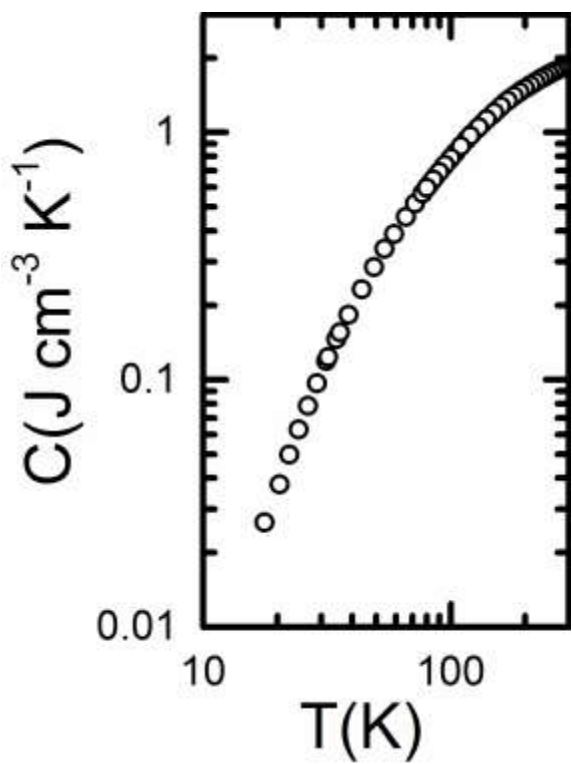

**Figure S12**. Volumetric heat capacity of black phosphorus from Ref. [19].

# SI References